\batchmode
\makeatletter
\def\input@path{{/Users/axelaraneda/Desktop/Research/fBM/}}
\makeatother
\documentclass[english]{article}
\usepackage{lmodern}
\usepackage[T1]{fontenc}
\usepackage[latin9]{inputenc}
\usepackage[a4paper]{geometry}
\usepackage{babel}
\usepackage{array}
\usepackage{varioref}
\usepackage{refstyle}
\usepackage{booktabs}
\usepackage{multirow}
\usepackage{amsmath}
\usepackage{amsthm}
\usepackage{amssymb}
\usepackage{graphicx}
\usepackage[numbers,numbers,sort&compress]{natbib}

\makeatletter

\AtBeginDocument{\providecommand\secref[1]{\ref{sec:#1}}}
\providecommand{\tabularnewline}{\\}
\RS@ifundefined{subsecref}
  {\newref{subsec}{name = \RSsectxt}}
  {}
\RS@ifundefined{thmref}
  {\def\RSthmtxt{theorem~}\newref{thm}{name = \RSthmtxt}}
  {}
\RS@ifundefined{lemref}
  {\def\RSlemtxt{lemma~}\newref{lem}{name = \RSlemtxt}}
  {}

\newcommand{\lyxaddress}[1]{
	\par {\raggedright #1
	\vspace{1.4em}
	\noindent\par}
}

\usepackage{hyperref}
\date{}

\usepackage{todonotes}
\usepackage{bbm}
\usepackage{bibentry}
\nobibliography*

\@ifundefined{showcaptionsetup}{}{%
 \PassOptionsToPackage{caption=false}{subfig}}
\usepackage{subfig}
\makeatother

\begin{document}
\title{\textbf{Credit Default Swaps }\linebreak{}
\textbf{and the mixed-fractional CEV model}\vspace{1em}}
\author{Axel A.~Araneda\thanks{Email: \protect\href{mailto:axelaraneda@mail.muni.cz}{\texttt{axelaraneda@mail.muni.cz}}} }
\maketitle

\lyxaddress{\begin{center}
\vspace{-2.5em} Institute of Financial Complex Systems \\Department
of Finance\\ Masaryk University\\ 602 00 Brno, Czech Republic.
\par\end{center}}

\begin{center}
\vspace{-1em} This version: \today \vspace{1.5em}
\par\end{center}
\begin{abstract}
This paper explores the capabilities of the Constant Elasticity of
Variance model driven by a mixed-fractional Brownian motion (mfCEV)
{[}\bibentry{araneda2020fractional}{]} to address default-related
financial problems, particularly the pricing of Credit Default Swaps.
The increase in both, the probability of default and the CDS spreads
under mixed-fractional diffusion compared to the standard Brownian
case, improves the lower empirical performance of the standard Constant
Elasticity of Variance model (CEV), yielding a more realistic model
for credit events. 


\vspace{2em}
\end{abstract}

\section{Introduction}

Credit Default Swaps (CDS) are derivatives designed to hedge default
in the firm's obligations, taking as reference some debt instrument
(e.g., bonds). It acts as insurance since the `protection buyer' is
compensated (usually with the debt face value) in case of a credit
default event in exchange for periodic payments to the `protection
seller'. On the other hand, when the reference asset is just equity
(stock price), and the triggering event is some pre-specified price-level
barrier (namely, 30\% or 50\% of the initial value), we are in front
of an Equity Default Swap (EDS), a deep out-of-the-money digital knock-in
put option, where the option premium, instead of paid at the inception,
is divided into legs along the contract duration to mimic the CDS
structure. In the case of a firm's debt default, the stock price would
be worth zero or very near to the zero level, then CDS are equivalent
to a zero-barrier EDS \citep{mendoza2011pricing}. 

From the price modelling perspective, the geometric Brownian motion,
the basic assumption of the seminal work of \citet{black1973pricing},
precludes hitting the zero price level and becoming ineligible to
address default-related problems. In that sense, the constant elasticity
of variance (CEV model) \citep{Cox1975,cox1996constant} emerges as
a candidate for this task due to some desirable properties. First,
the origin is an attainable and absorbing boundary\footnote{\label{fn:The-attainable-boundary}The attainable boundary at zero
occurs (a.s) when the elasticity of variance is negative or in our
notation (c.f \secref{The-mixed-fractional-CEV}) $\alpha<2$. The
absorbing condition at zero is naturally given for $\alpha\leq1$;
while for a $1<\alpha<2$ the origin could be absorbing or reflecting.
For financial purposes the absorbing condition at zero is imposed:
when the price reaches zero the diffusion process is killed (cemetery
state). Please see \citep{lindsay2012simulation} for a detailed analysis
of boundary conditions in the CEV model. } via diffusion to zero, and consequently, it includes the bankruptcy
possibility. Second, the CEV model addresses some empirical facts
observed in equity markets: the leverage effect, heteroskedasticity,
and the implied volatility skew.

The standard CEV assumption has already been applied to credit risk
pricing \citep{campi2005closed,albanese2005pricing}. However, the
lower CDS spreads under the standard CEV model compared to the actual
market data \citep{carr2006jump,mendoza2011pricing}, yields to consider
some extensions as the addition of a Heston-type stochastic volatility
feature to the CEV stock price \citep{atlan2005hybrid}, or the inclusion
of a killing jump rate \citep{campi2009systematic,carr2006jump,mendoza2011pricing}.

In this note, we want to enrich the CEV-CDS literature, including
a non-standard diffusion mechanism for the CEV model, particularly
the mixed-fractional Brownian motion (mfBm) which provides to the
CEV model the capacity to address the long-range dependence empirical
observation without sacrificing the martingale property and a better
fit with market option prices \citep{araneda2020fractional,ARANEDA2021125974}.

\section{The mixed-fractional CEV model  \label{sec:The-mixed-fractional-CEV}}

The mfCEV establishes the following stochastic differential equation
for the evolution of the asset price $S$ \citep{araneda2020fractional}:

\begin{equation}
\mathrm{d}S_{t}=rS_{t}\mathrm{d}t+\delta S_{t}^{\frac{\alpha}{2}}\mathrm{d}M_{t}^{H,\beta}\label{eq:SCEV}
\end{equation}
\noindent where $M_{t}^{H}$ is a mfBm defined as:

\begin{equation}
M_{t}^{H,\beta}=B_{t}+\beta B_{t}^{H}\label{eq:mF}
\end{equation}

\noindent being $\beta\in\mathbb{R}_{0}^{+}$, $B_{t}^{H}$ a fractional
Brownian motion with Hurst parameter $\frac{3}{4}<H<1$, and $B_{t}$
an independent $\&$ standard Bm. These conditions guarantee that
the mfBm is a local-martingale equivalent in law to a standard Brownian
motion \citep{cheridito2001mixed}. The parameter $\alpha$ is restricted
to less than 2 in order to ensure three features: i) zero is an attainable
and absorbing boundary (cf. footnote \ref{fn:The-attainable-boundary});
ii) inverse relationship between price and volatility (Leverage effect);
and iii) arbitrage-free (which is possible because $M_{t}^{H,\beta}$
is semi-martingale for $H>3/4$; see \citep{lindsay2012simulation,delbaen2002note}
about the CEV model and arbitrage possibilities). The other two parameters
of the model, $r$ and $\delta$, are assumed non-negative and strictly
positive, respectively. The former represents the constant risk-free
interest rate while the latter adjusts the at-the-money volatility
level and is usually parametrized as\label{parametrization} $\delta^{2}=\sigma_{0}^{2}S_{0}^{2-\alpha}$,
so $\sigma_{0}$ acts as the volatility scale parameter of the local
volatility function; i.e., $\sigma\left(S,0\right)=\sigma_{0}$.

Defining the new variable $x_{t}=S_{t}^{2-\alpha}$, the related Wick-It\^o
calculus (Appendix \ref{sec:It-Wick-formula-for}) yields to :

\begin{eqnarray*}
\text{d}x_{t} & = & \left(2-\alpha\right)\left[rx_{t}+\sigma^{2}\left(1-\alpha\right)\left(\frac{1}{2}+\beta^{2}Ht^{2H-1}\right)\right]\text{d}t+\left(2-\alpha\right)\sigma\sqrt{x_{t}}\text{d}M_{t}^{H,\beta}
\end{eqnarray*}

\noindent and the corresponding transition density probability function
$P=P\left(\left.x_{t},t\right|x_{0},0\right)$ obeys (c.f. Appendix
\ref{sec:Effective-Fokker-Planck-equation}):

\begin{equation}
\frac{\partial P}{\partial t}=-\frac{\partial}{\partial x_{t}}\left\{ \left[Ax_{t}+B\left(t\right)\right]P\right\} +C\left(t\right)\frac{\partial^{2}}{\partial x_{t}^{2}}\left[xP\right]\label{eq:FPx}
\end{equation}

\noindent with $A=\left(2-\alpha\right)r,$ $B\left(t\right)=\sigma^{2}\left(1-\alpha\right)\left(2-\alpha\right)\left(\frac{1}{2}+\beta^{2}Ht^{2H-1}\right)$,
and $C\left(t\right)=\sigma^{2}\left(2-\alpha\right)^{2}\left(\frac{1}{2}+\beta^{2}Ht^{2H-1}\right)$. 

Eq. (\ref{eq:FPx}) is equivalent to the Fokker-Planck equation for
a time-inhomogeneous Feller (square-root) process, and given the ratio
$\theta=B(t)/C(t)=\left(1-\alpha\right)/\left(2-\alpha\right)<1$
is time-independent\footnote{This ensures that origin is an attanable boundary according to the
Feller classification in the defined domain for $\alpha$; i.e., regular
boundary for $\alpha<1$ and absorving boundary for $1\leq\alpha<2$}, it could be solved analytically \citep{araneda2020fractional,giorno2021time}.
Moreover, considering these conditions, the density $g$ of the random
variable $\tau=\inf\left\{ t>0:x_{t}=0\right\} $ which describes
first-passage time (FPT) through the zero state is given by\footnote{In their paper \citet{giorno2021time} imposse the condition $0\le\xi<1$;
however the supplied results are still valid for negative $\xi$ ratios.} \citep{giorno2021time}:

\begin{eqnarray}
g\left(\left.0,t\right|x_{0},0\right) & = & \frac{1}{\Gamma\left(1-\xi\right)}\frac{\sigma^{2}\left(2-\alpha\right)^{2}\text{e}^{-\left(2-\alpha\right)rt}}{\phi\left(t\right)}\left[\frac{x_{0}}{\phi\left(t\right)}\right]^{1-\xi}\exp\left[-\frac{x_{0}}{\phi\left(t\right)}\right]\nonumber \\
 & = & -\frac{1}{\Gamma\left(1-\xi\right)}\frac{\partial}{\partial t}\gamma\left(1-\xi,{\displaystyle \frac{x_{0}}{\phi\left(t\right)}}\right)\label{eq:g2}
\end{eqnarray}

\noindent being $\gamma\left(\bullet,\bullet\right)$ the lower incomplete
gamma function, and 

\begin{multline*}
\phi\left(t\right)=\int_{0}^{t}C\left(\tilde{t}\right)\text{e}^{-\left(2-\alpha\right)r\tilde{t}}\text{d}\tilde{t}=\frac{\sigma^{2}\left(2-\alpha\right)}{2r}\left[1-\text{e}^{-\left(2-\alpha\right)rt}\right]\\
+\beta^{2}\frac{\sigma^{2}\left(2-\alpha\right)^{2}}{2\left(2H+1\right)}\text{e}^{-\left(2-\alpha\right)rt}t^{2H}\left\{ 2H+1+\text{e}^{\frac{1}{2}\left(2-\alpha\right)rt}\left[\left(2-\alpha\right)rt\right]^{-H}M_{H,H+1/2}\left[\left(2-\alpha\right)rt\right]\right\} 
\end{multline*}

\noindent with $M_{\kappa,\upsilon}\left(l\right)$ the M-Whittaker
function.

Thus, the first-passage-time (FPT) probability ($Q$) is equal to
\citep{giorno2021time}:

\[
Q\left(\left.\tau\leq t\right|x_{0},0\right)=1-\frac{\gamma\left(1-\xi,\frac{x_{0}}{\phi\left(t\right)}\right)}{\Gamma\left(1-\xi\right)}=\frac{\Gamma\left(1-\xi,{\displaystyle \frac{x_{0}}{\phi\left(t\right)}}\right)}{\Gamma\left(1-\xi\right)}
\]

\noindent where $\Gamma\left(\bullet,\bullet\right)$ denotes the
upper incomplete gamma function, and in consequence, the risk-neutral
default probability for the mfCEV model is expressed as:

\[
Q\left(\left.\tau\leq t\right|S_{0},0\right)=\frac{\Gamma\left(1-\xi,{\displaystyle \frac{S_{0}^{2-\alpha}}{\phi\left(t\right)}}\right)}{\Gamma\left(1-\xi\right)}
\]

\begin{figure}
\subfloat[CEV\label{fig:blue}]%
{\includegraphics[width=0.5\linewidth]{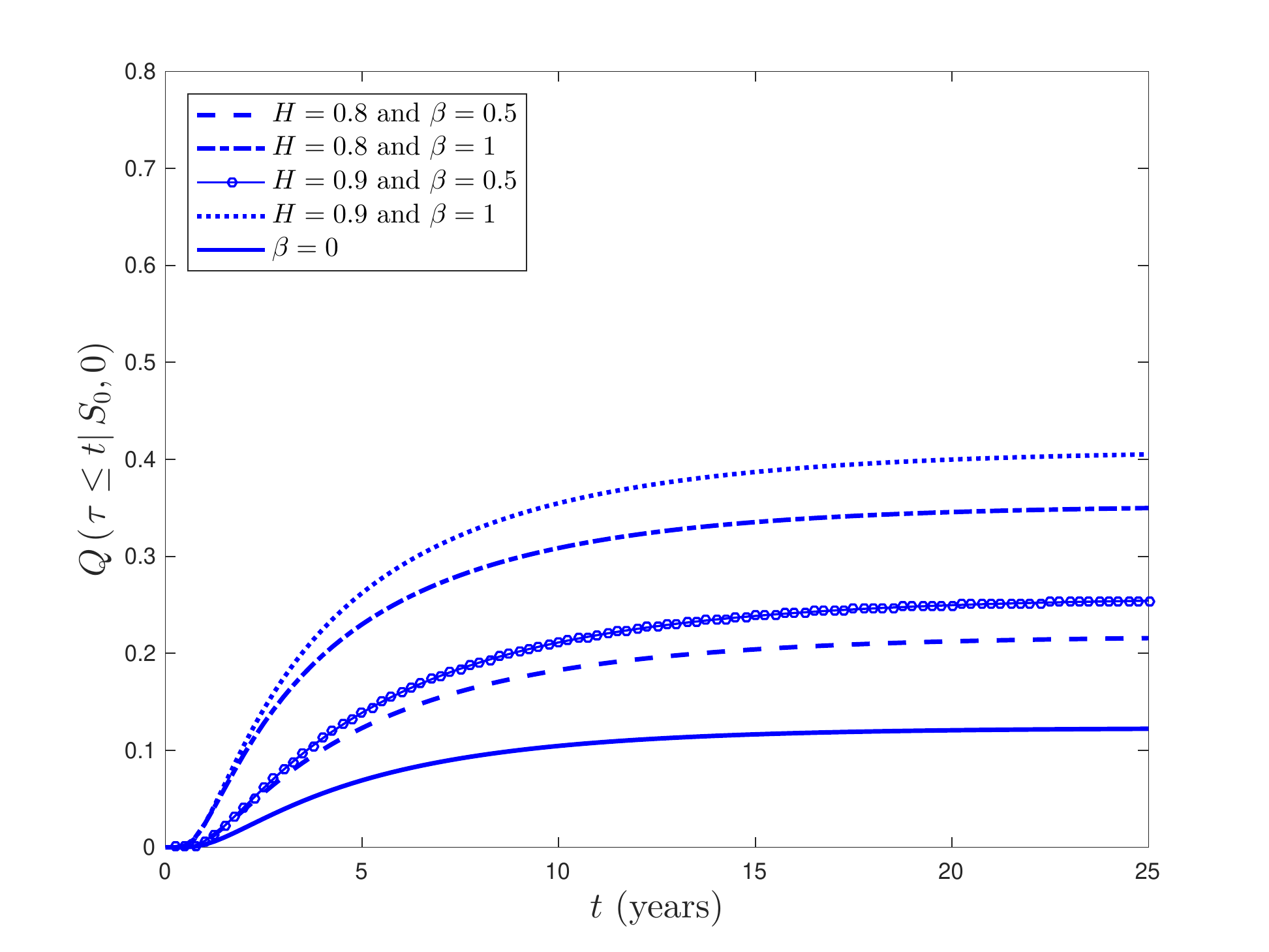}

}%
\subfloat[$\alpha=0$ and $\sigma_{0}=20\%$\label{fig:red}]%
{\includegraphics[width=0.5\linewidth]{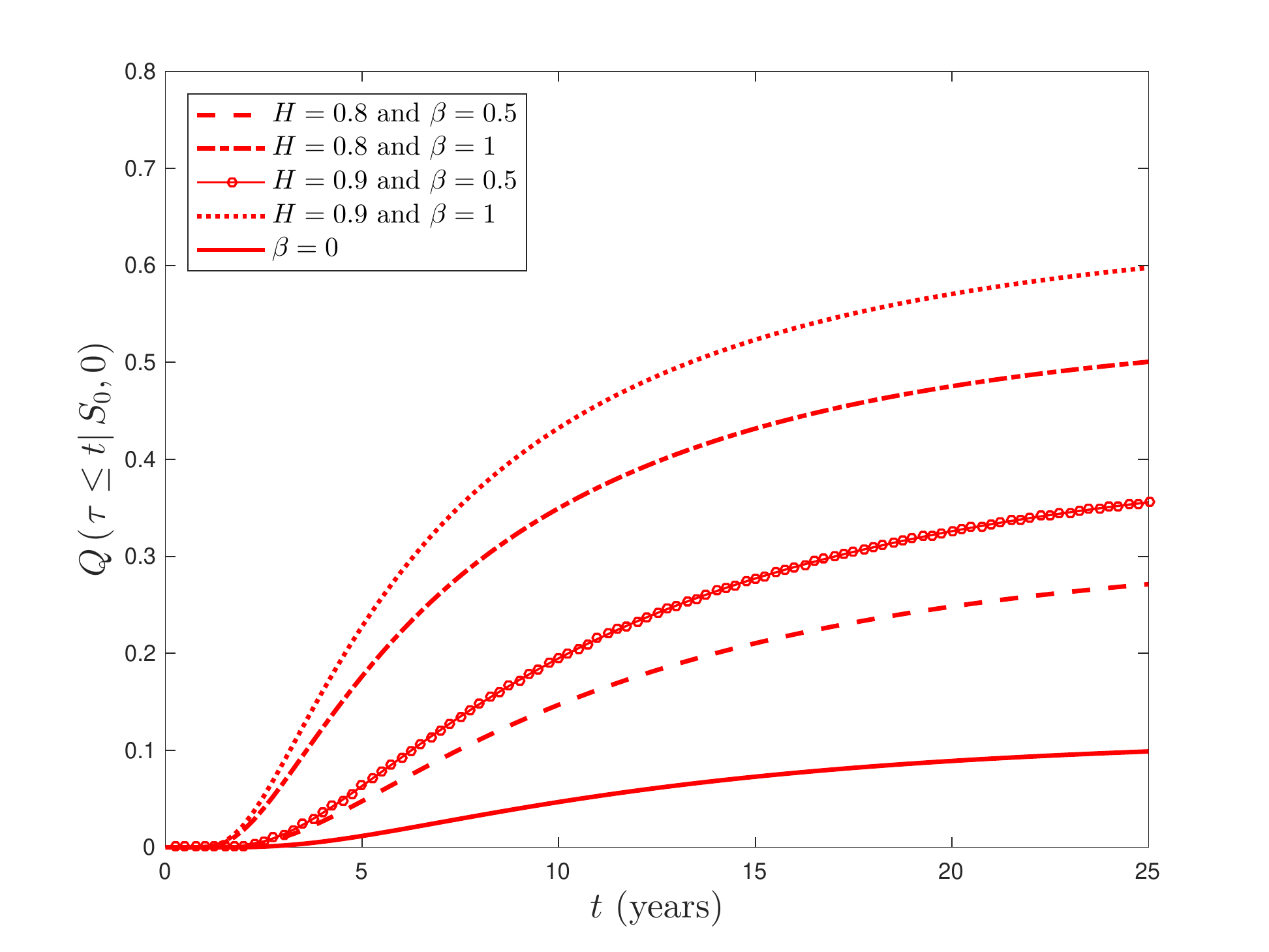}

}

\caption{Probability of default (price equal to zero) under both CEV and mfCEV
models, setting $S_{0}=50$, $\sigma_{0}=20\%$ and $r=5\%$.\label{fig:Probability-of-default}}

\end{figure}

Fig. \ref{fig:Probability-of-default} plots the FPT probability through
the zero state as a function of time for the mixed-fractional CEV
model under different elasticity parameters: $\alpha=-2$ (blue, \ref{fig:blue})
and $\alpha=$0 (red, \ref{fig:red}). Moreover, to visualize the
mfCEV model sensitivity to both the Hurst exponent and $\beta$, we
use $H=\left\{ 0.8,0.9\right\} $ and $\beta=\left\{ 0,1/2,1\right\} $.
The case $\beta=0$ corresponds to the standard CEV specification.
We have set the initial volatility\footnote{The parametrization given \vpageref{parametrization} for the local
volatility function makes the defaul probability (and the CDS pricing
addressed in the following section) independent of the initial price
level.} $(\sigma_{0})$ at 20\%, and the risk-free rate equals 5\%. We see
the mixed-fractional diffusion increases the default probability (as
a function of both $\beta$ and $H$) compared to the classical CEV.
The increment in the FPT is faster for a greater $\alpha$ at long
times. Nevertheless, for short maturities, the increment is more pronounced
under lower $\alpha$. 

\section{CDS rates}

The objective here is to compute both i) the present value ($V$)
of the protection payment to the buyer due to the triggering event
(100\%-drop in the price level) and ii) the swap rate. For a maturity
$T$, a constant risk-free rate $r$, a notional amount of 1\$, and
a fixed recovery rate $R$, the risk-neutral valuation yields: 

\[
V=\left(1-R\right)\cdot\mathbb{E}\left(\text{e}^{-r\tau}\mathbbm{1}_{\tau<T}\right)
\]

\noindent where the indicator function $\mathbbm{1}_{\tau<T}$ maps
a price default event in the time-interval $[0,T]$. Then,

\begin{eqnarray}
V & = & \left(1-R\right)\int_{0}^{T}\text{e}^{-rt}g\left(\left.0,t\right|x_{0},0\right)\text{d}t\nonumber \\
 & = & \left(1-R\right)\left[\text{e}^{-rT}Q\left(\left.\tau\leq t\right|S_{0},0\right)+r\int_{0}^{T}\text{e}^{-rt}Q\left(\left.\tau\leq t\right|S_{0},0\right)\right]\label{eq:Vg}
\end{eqnarray}

On the other hand, in CDS contracts, the protection payment is divided
in installments from the inception time up to either the trigger event
or the expiration time, whichever comes first.  Then, ignoring any
accrual payment after the default, the equilibrium swap rate (coupon
$C$) under typical conditions (semiannual installments and a maturity
equal to an integer number of years) is given by:

\[
C={\displaystyle \frac{V}{{\displaystyle \sum_{i=1}^{2T}}\text{e}^{-r\frac{i}{2}}\mathbb{E}\left[\mathbbm{1}_{\tau>\frac{i}{2}}\right]}=}{\displaystyle \frac{V}{{\displaystyle \sum_{i=1}^{2T}}\text{e}^{-r\frac{i}{2}}\left[1-Q\left(\left.\tau\leq\frac{i}{2}\right|S_{0},0\right)\right]}}
\]

Table (\ref{tab:Benchmark-EDS-fee}) shows the CDS spreads, i.e.,
the annual coupon payments (in basis points), when the underlying
asset follows a mfCEV ($\beta=\left\{ 0.5,1\right\} $, $H=\left\{ 0.8,0.9\right\} $)
and a standard CEV ($\beta=0$), with $\sigma_{0}=20\%$, $r=5\%$,
a recovery rate of 50\%, $\alpha=\left\{ 0,-2\right\} $; under different
maturities, being $V$ computed by numerical integration of Eq. (\ref{eq:Vg}).
The results confirm the dependence on both maturity and elasticity;
and the lower swap rate for the classical CEV for low maturities,
due to the low default probability under the standard Brownian diffusion.
Meanwhile, the mfCEV provides larger coupon values, even at low tenors,
as a function of both $H$ and $\beta$, allowing a flexible calibration
to match the term structure of CDS spreads.

\begin{table}
\begin{centering}
\par\end{centering}
\begin{centering}
\begin{tabular}{cccccccccc}
\toprule 
 &
 &
\multicolumn{2}{c}{$T=1$} &
\multicolumn{2}{c}{$T=2$} &
\multicolumn{2}{c}{$T=5$} &
\multicolumn{2}{c}{$T=10$}\tabularnewline
\midrule 
 &
 &
$\alpha=0$ &
$\alpha=-2$ &
$\alpha=0$ &
$\alpha=-2$ &
$\alpha=0$ &
$\alpha=-2$ &
$\alpha=0$ &
$\alpha=-2$\tabularnewline
\midrule
\midrule 
$\beta=0$ &
 &
0.0015 &
14.6761 &
0.4976 &
49.3693 &
11.0929 &
71.0707 &
22.0907 &
58.1472\tabularnewline
\midrule 
\multirow{2}{*}{$\beta=0$.5} &
$H=0.8$ &
0.0220 &
33.0638 &
3.6859 &
97.5923 &
45.8409 &
130.6805 &
73.6537 &
107.2735\tabularnewline
\cmidrule{2-10} \cmidrule{3-10} \cmidrule{4-10} \cmidrule{5-10} \cmidrule{6-10} \cmidrule{7-10} \cmidrule{8-10} \cmidrule{9-10} \cmidrule{10-10} 
 & $H=0.9$ &
0.0219 &
32.9327 &
4.5121 &
104.3824 &
61.1677 &
148.0252 &
99.5110 &
125.2780\tabularnewline
\midrule 
\multirow{2}{*}{$\beta=1$} &
$H=0.8$ &
1.3802 &
121.9533 &
45.2696 &
250.5198 &
182.4174 &
265.8567 &
206.8295 &
206.2857\tabularnewline
\cmidrule{2-10} \cmidrule{3-10} \cmidrule{4-10} \cmidrule{5-10} \cmidrule{6-10} \cmidrule{7-10} \cmidrule{8-10} \cmidrule{9-10} \cmidrule{10-10} 
 & $H=0.9$ &
1.3665 &
121.0740 &
58.1627 &
275.5237 &
240.6370 &
307.8064 &
270.6823 &
244.4577\tabularnewline
\bottomrule
\end{tabular}
\par\end{centering}
\caption{CDS spread under both standard and mixed-fractional CEV model\label{tab:Benchmark-EDS-fee}}
\end{table}

\bibliographystyle{unsrtnat}
\bibliography{14_Users_axelaraneda_Desktop_Research_fBM_fBM}

\appendix

\section{It\^o-Wick formula for mfBm\label{sec:It-Wick-formula-for}}

From Eq. (\ref{eq:mF}), we have that:

\[
\mathbb{E}\left[\left(M_{t}^{H,\beta}\right)^{2}\right]=t+\beta^{2}t^{2H}
\]

Given that the variance is bounded, for $f=f\left(t,x\left(M_{t}^{H,\beta}\right)\right)\in C^{1,2}\left(\left(0,\infty\right)\times\mathbb{R}\right)$
we have \citep{nualart2008wick}:

\begin{eqnarray}
\text{d}f & = & \frac{\partial f}{\partial t}\text{d}t+\frac{\partial f}{\partial x}\text{d}M_{t}^{H,\beta}+\frac{1}{2}\text{d}\left(t+\beta^{2}t^{2H}\right)\frac{\partial^{2}f}{\partial x^{2}}\nonumber \\
 &  & =\frac{\partial f}{\partial t}\text{d}t+\frac{\partial f}{\partial x}\text{d}M_{t}^{H,\beta}+\frac{1}{2}\left(1+2H\beta^{2}t^{2H-1}\right)\text{d}t\frac{\partial^{2}f}{\partial x^{2}}\nonumber \\
 &  & =\left[\frac{\partial f}{\text{\ensuremath{\partial t}}}+\left(\frac{1}{2}+H\beta^{2}t^{2H-1}\right)\frac{\partial^{2}f}{\text{\ensuremath{\partial x^{2}}}}\right]\text{d}t+\frac{\partial f}{\text{\ensuremath{\partial x}}}\text{d}M_{t}^{H,\beta}\label{eq:Ito}
\end{eqnarray}

\section{Effective Fokker-Planck equation\label{sec:Effective-Fokker-Planck-equation}}

Let $y_{t}$ a stochastic process described by the following stochastic
differential equation (SDE) driven by a mfBm:

\begin{equation}
\text{d}y_{t}=\mu\left(y_{t},t\right)\text{d}t+\sigma\left(y_{t},t\right)\text{d}M_{t}^{H,\beta}\label{eq:dy}
\end{equation}

\noindent then, the transformation $g=g\left(y_{t},t\right)\in C^{2}\left(\mathbb{R}\right)$
follows:

\begin{equation}
\text{d}g\left(y_{t}\right)=\left[\left(\frac{\partial g}{\text{\ensuremath{\partial y}}}\right)\mu+\left(\frac{1}{2}+H\beta^{2}t^{2H-1}\right)\sigma^{2}\frac{\partial^{2}g}{\text{\ensuremath{\partial y^{2}}}}\right]\text{d}t+\left(\frac{\partial g}{\text{\ensuremath{\partial y}}}\right)\sigma\text{d}M_{t}^{H,a,b}\label{eq:dg}
\end{equation}

Taking expectations at both sides of Eq. (\ref{eq:dg}) and using
that $\mathbb{E}\left[g\left(y_{t},t\right)\right]=\int g\cdot P\left(y_{t},t\right)$,
where $P=P\left(y_{t},t\right)$ is the transition density of the
random variable $y$ at time $t$; we got:

\begin{equation}
\int_{-\infty}^{\infty}g\frac{\partial P}{\partial t}\text{d}y=\int_{-\infty}^{\infty}\left[\left(\frac{\partial g}{\text{\ensuremath{\partial y}}}\right)\mu+\left(\frac{1}{2}+H\beta^{2}t^{2H-1}\right)\sigma^{2}\frac{\partial^{2}g}{\text{\ensuremath{\partial y^{2}}}}\right]P\text{d}y\label{eq:FPg}
\end{equation}

Taking in account that $\int_{-\infty}^{\infty}\left(\frac{\partial g}{\text{\ensuremath{\partial y}}}\right)\mu P\text{d}y=-\int_{-\infty}^{\infty}g\frac{\partial\left(\mu P\right)}{\partial y}\text{d}y$
and $\int_{-\infty}^{\infty}\sigma^{2}\frac{\partial^{2}g}{\text{\ensuremath{\partial y^{2}}}}P\text{d}y=-\int_{-\infty}^{\infty}g\frac{\partial^{2}\left(\sigma^{2}P\right)}{\partial y^{2}}\text{d}y$,
the transition density for the variable $y_{t}$ described by Eq.
(\ref{eq:dy}) is ruled by:

\begin{equation}
\frac{\partial P}{\text{\ensuremath{\partial t}}}=-\frac{\partial\left(\mu P\right)}{\partial y}+\left(\frac{1}{2}+H\beta^{2}t^{2H-1}\right)\frac{\partial^{2}\left(\sigma^{2}P\right)}{\partial y^{2}}\label{eq:FP}
\end{equation}

\end{document}